\documentclass[12pt]{article}
\usepackage{aas_macros}
\usepackage{indentfirst}
\usepackage[utf8]{inputenc}
\usepackage{pifont}
\usepackage{graphicx}
\usepackage{times}
\usepackage{enumitem,amssymb}
\usepackage{ragged2e}
\usepackage{natbib}
\usepackage{titling}
\usepackage{setspace}
\usepackage{changes}
\definechangesauthor[color=red,name={Joe Lazio}]{JL}
\usepackage[bookmarks=false]{hyperref}
\RequirePackage{doi}
\hypersetup{colorlinks=true,citecolor=blue,linkcolor=blue,urlcolor=blue,breaklinks=true}
\begin{document}
\bibliographystyle{plainnat}
\setcounter{page}{0}

\begin{titlepage}
\setlength{\parindent}{0pt}
            {\huge Astro2020 Science White Paper \par}        
	\vspace{1cm}
               {\huge\bfseries The radio search for technosignatures\\in the decade 2020--2030\par}
	\vspace{1cm}
	\includegraphics[width=6.5in]{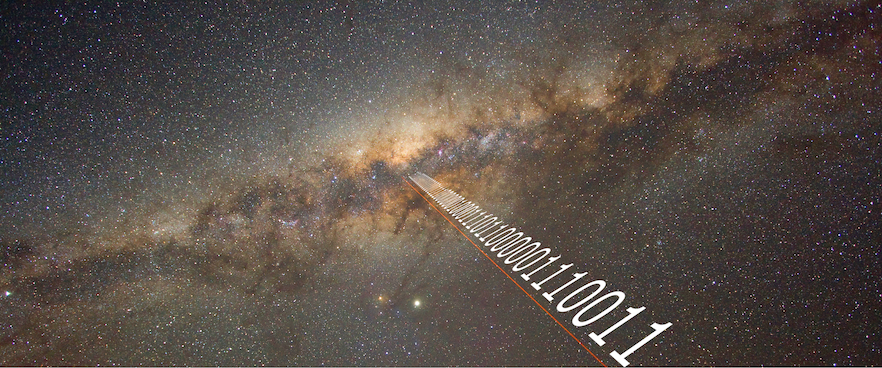}\par
        \vspace{0.1cm}
\centering
          {\scriptsize Background photo: central region of the Galaxy by Yuri Beletsky, Carnegie Las Campanas Observatory}\par
          \justifying
          \setlength{\parindent}{0pt}
	  \vspace{1cm}
                    {{\bf Thematic Area:} \par Planetary Systems\par}
	  \vspace{1cm}
               {\bfseries Principal Author: }\par
               \vspace{0.2cm}
               { Name: Jean-Luc Margot}\par
	       \vspace{0.2cm}
               { Institution: University of California, Los Angeles}\par
	       \vspace{0.2cm}
               { Email: jlm@astro.ucla.edu}\par
	       \vspace{0.2cm}
               { Phone: 310.206.8345}\par
	       \vspace{0.4cm}
	\vfill
               {\bfseries Co-authors: }\par
               \vspace{0.2cm}
               Steve Croft (University of California, Berkeley), T.\ Joseph W.\ Lazio (Jet Propulsion Laboratory), Jill Tarter (SETI Institute), Eric J.\ Korpela (University of California, Berkeley)
               
	\vfill

	{\today\par}
\end{titlepage}

\newpage

\setcounter{page}{1}
\section{Scientific context} \label{sec:intro}
Are we alone in the universe?  This question is one of the most profound scientific questions of our time.  All life on Earth is related to a common ancestor, and the discovery of other forms of life will revolutionize our understanding of living systems.  On a more philosophical level, it will transform our perception of humanity's place in the cosmos.

Observations with the NASA Kepler telescope have shown that there are billions of habitable worlds in our Galaxy~\citep[e.g.,][]{boru16}. The profusion of planets, coupled with the abundance of life's building blocks in the universe, suggests that life itself may be abundant.

Currently, the two primary strategies for the search for life in the universe are (1) searching for biosignatures in the Solar System or around nearby stars and (2) searching for technosignatures emitted from sources in the Galaxy and beyond~\citep[e.g.,][]{nas18life}.  Given our present knowledge of astrobiology, there is no compelling reason to believe that one strategy is more likely to succeed than the other.  Advancing the scientific frontier in this area therefore requires support of  {\em both} strategies~\citep[e.g.,][]{nas90life}.
In addition, the search for technosignatures is logically continuous with the search for biosignatures: If one searches for signatures of biology, including single-celled organisms, it stands to reason to also search for signatures of complex life (multi-cellular organisms), including technology.
The recent report from the NASA technosignatures workshop~\citep{technosigs18} suggests that the search for technosignatures %
is not only complementary to the search for biosignatures but also provides some advantages:
\begin{quote}
  Compared to biosignatures, technosignatures might therefore be more ubiquitous, more obvious, more unambiguous, and detectable at much greater ... distances.
\end{quote}

We expand on this statement and identify four areas of strength that pertain to the search for radio technosignatures:
\begin{enumerate}

  \item {\bf Cost.}  One can implement a search for technosignatures
    that
    dramatically expands the search volume sampled to date for
    $\sim$\$10
    million per year.  The annual budget for NASA's SETI program at
    the time of its cancellation in 1993 was $\sim$\$10
    million \citep{garb99}.  The Breakthrough Listen (BL) project
    aspires to fund a search at a similar level~\citep{word17}.
    In
    contrast, cost projections for Mars and Europa landers are $\sim$billions of dollars, and the expected cost of the James
    Webb Space Telescope is $\sim$\$10 billion.  {\bf A modest budgetary increment can
      expand the search for life in the universe from primitive to complex life and from the solar neighborhood to the entire Galaxy.}
\item {\bf Search volume.}
  Current technology enables the detection of technosignatures emitted thousands of light years away \citep[e.g.,][]{marg18}.  In contrast, robotic landers can sample only a few nearby planets and satellites.  Likewise, telescopic searches for biosignatures are limited to a modest number of planetary systems
  in a small region ($\lesssim$50 light years) around the Sun.
  The volume of the Galaxy that can be sampled with a radio search for technosignatures is millions of times larger than the relatively small, local bubble conducive to the search for biosignatures.  {\bf The number of accessible targets in the search for technosignatures exceeds the number of accessible targets in the search for biosignatures by a factor of a billion or more, a situation that will persist in the next decade.}

\item {\bf Certainty of interpretation.}  Although some types of biosignatures (e.g., a fossil or sample organism) may offer compelling interpretations, many of the proposed exoplanet biosignatures are expected to be difficult to detect and to yield inconclusive interpretations at least until 2030 \citep[e.g.,][]{fuji18}.  %
  Confusion with abiogenic sources may remain difficult to rule out~\citep[e.g.,][]{rein14}.  In many cases, the spectroscopic observations may be consistent with, but not diagnostic of, the presence of life~\citep{catl18}.  In contrast, the search for technosignatures provides the opportunity {\em today} to make robust detections with irrefutable interpretations.
   If repeatable, narrowband or artificially pulsed emissions from an extraterrestrial emitter are detected, the interpretation will be
  unambiguous and compelling.

\item {\bf Potential information content.}
  The information content of a telescopic detection
    of strong thermochemical disequilibrium in an exoplanet atmosphere
     will remain rather limited.
    We are unlikely to
    infer, for instance, whether the putative organisms are primitive
    or evolved.  A fossil or sample organism recovered from Mars or an
    ocean world in the Solar System would provide orders of magnitude
    more information (the human genome encodes
    approximately $3 \times 10^9$ bits).  Detection of a radio
    technosignature would immediately indicate that another evolved
    life form existed at some time in the Galaxy.
    The detection may or may not provide additional information, depending on the modulation of the signal, but the potential for
    profound advances in human knowledge exists.
    A simple
    binary code with a bandwidth of 100~Hz -- comparable to 19th
    century telegraphy -- can deliver $3 \times 10^9$ bits per year.
    The size of the article text in the English Wikipedia is about
    $10^{11}$ bits (compressed).

\end{enumerate}

The scientific context for the radio search for technosignatures includes a nearly 60-year legacy~\citep[][]{cocc59,drak61} and has been well reviewed~\citep[e.g.][]{tart01,eker02,tart10}.  Since these reviews and the Astro2010 decadal survey, the paradigm for the search for technosignatures has changed dramatically.  The newly acquired knowledge that most stars have planets and that a substantial fraction of stars (conservatively at least 10\%) host habitable worlds has amplified the motivation to search.

\section{Scientific opportunities}\label{sec:conclusions}

  \subsection{Fraction of search volume sampled so far}
It is useful to consider the fraction of the search volume, also known as ``Cosmic Haystack'' \citep{tart10}, that has been sampled so far.  
The dimensions of the search volume
\citep[e.g.,][]{drak84,wright18haystack}
normally include sensitivity, sky coverage, frequency coverage, and time.  \citet{tart10} estimated that we have sampled between $5\times10^{-26}$ and $1.5\times10^{-18}$ of the 9-dimensional cosmic haystack, where the range of values reflects the longevity of the transmitter, assumed to be between 100 years and 3 billion years.  Other assumptions in the calculation include $10^{11}$ stars to search, 9 billion 1-Hz channels within the 1--10 GHz radio frequency range, 100 different modulation schemes, 100 different dispersion values, and detection of an Arecibo-class transmitter.
With Tarter's analogy, we have sampled between 0.07 ml and 2100 l out of the Earth's oceans, or a geometric mean of 0.4 l (1.6 US cups of water).  One can choose slightly different values in the calculation,
but the basic result remains: the fraction of the search volume sampled to date is minute, and there is a tremendous opportunity to expand the search.  

    \subsection{Expansion of the sampled fraction in the next decade}
    The most useful search programs can detect radio transmissions that are either intended for detection by a distant civilization (e.g., a beacon) or not (e.g., a radar or inter-planet telecommunication system).  Because the transmitter luminosity function is unknown, it is difficult to optimize search strategies \citep[e.g.,][]{eker02}, and a variety of approaches are warranted.  
    Both targeted surveys and broad surveys are required to examine all dimensions of the cosmic haystack.  Targeted surveys are necessary to sample regions at high sensitivity, whereas broad surveys are ideal to maximize sky coverage.  In both cases, efforts should be made to maximize the frequency coverage and time span of observations (``all frequencies, all the time'').  Sensitivity is a particularly important dimension of the cosmic haystack because the spatial volume from which transmitting technological species can be detected scales as the minimum detectable flux to the (-3/2) power.
    For instance, the search volumes of the next generation Very Large Array (ngVLA) and Square Kilometer Array (SKA) are expected to be $\sim$20 and $\sim$500 times larger than those of existing 100-m class telescopes, respectively (Figure 1).
    Searches for technosignatures in this decade have surveyed hundreds \citep[e.g.,][]{enri17} to thousands \citep[e.g.,][]{harp16} of stars.  Recent development in hardware and software capabilities, including beamforming with arrays of antennas,
    will enable an increase in the size of the sample by at least two orders of magnitude in 2020--2030.

\begin{figure}[h!]
	\includegraphics[width=6.5in,clip]{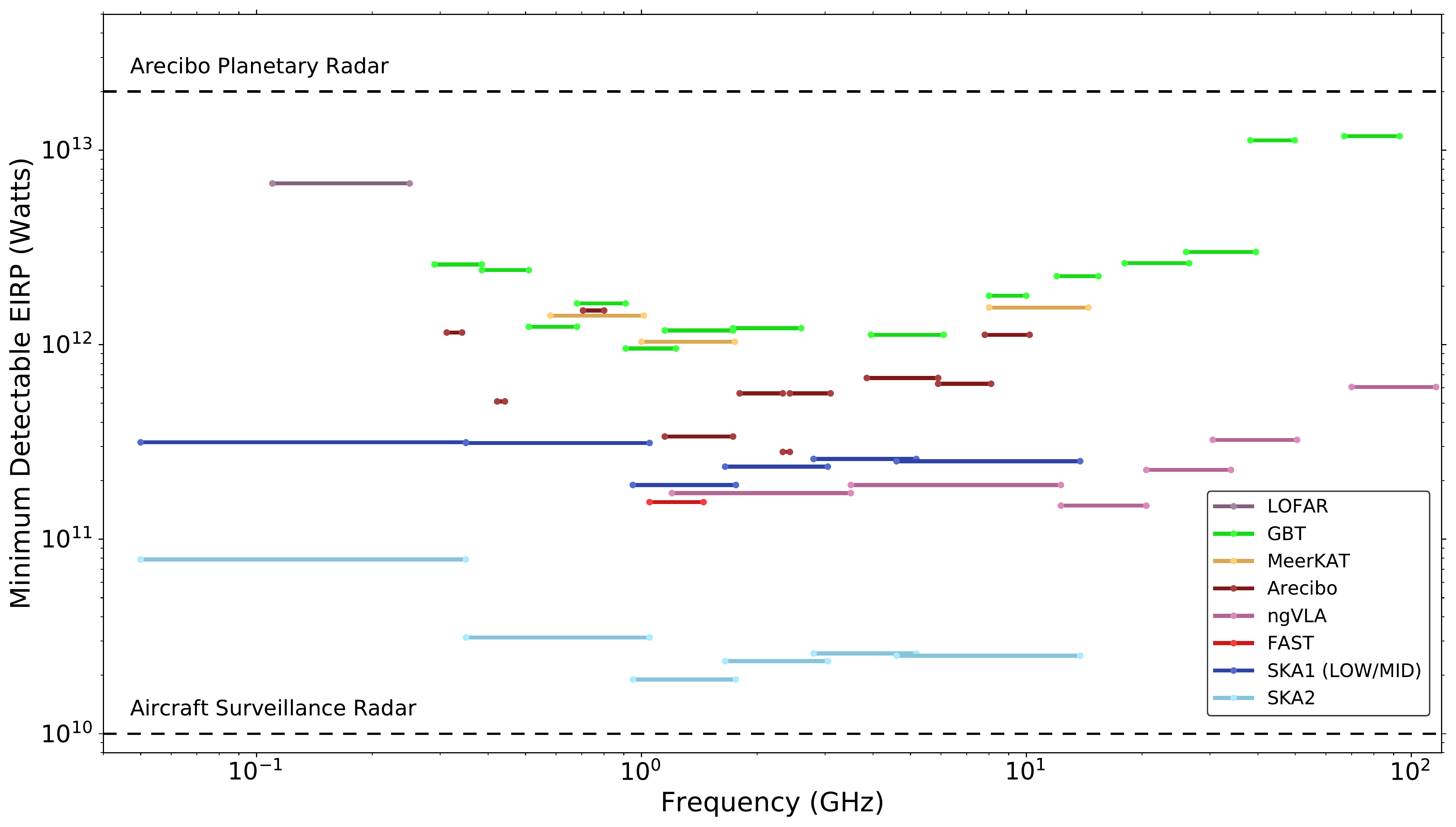}\par
        \caption{Sensitivity of existing and planned telescopes to narrowband signals emitted from a source located 100 light years away, adapted from
          \citet{siem13} and \citet{crof18}.  
          A transmitter is detectable if its equivalent isotropic radiated power (EIRP) is above the curve for a given telescope.
          Assumptions include a detection signal-to-noise ratio of 10, bandwidth of 3~Hz, and integration time of 5 minutes.
          With these assumptions, most facilities can detect Arecibo's EIRP from $\sim$500 light years away and 1000 $\times$ Arecibo's EIRP from $\sim$16,000 light years away.
        }
          
\end{figure}
    
    \subsection{Opportunities with commensal observing time}
    Because the search volume is so vast, it is important to maximize telescope access and to conduct technosignature searches while radio telescopes are being used for other scientific investigations.  This commensal mode of observing has been used successfully by, e.g., the SERENDIP \citep{wert97}, SETI@home \citep{korp12}, and ATA projects.  In commensal mode, the primary observer dictates the pointing and cadence of observations, which prevents the secondary observer from using an optimal cadence (e.g., to mitigate radio-frequency interference (RFI))
    and/or re-observation strategy (e.g., to confirm the near-real-time detection of interesting candidate signals).  The development of telescope arrays with beamforming capability largely alleviates this disadvantage by allowing the secondary observer to electronically point the telescope within the primary beam chosen by the primary observer.  The Breakthrough Listen (BL) team intends to use this capability on the
    Square Kilometer Array
    SKA precursor MeerKAT to survey $\sim$750,000 stars per year and enough sensitivity to detect an Arecibo-class transmitter, i.e., equivalent isotropic radiated power (EIRP) of $2\times10^{13}$ W, out to $\sim$700 light years.

    \subsection{Opportunities with dedicated observing time}
    \label{sec-dedicated}
    The penalties associated with commensal observations emphasize the need for telescope time dedicated to the search for technosignatures.      Existing instruments (e.g., Green Bank Telescope, Parkes Radio Telescope, Arecibo Observatory, NASA/JPL Deep Space Network,
    Five-hundred-meter Aperture Spherical radio Telescope, MeerKAT, the Murchison Widefield Array, the Low-Frequency Array, the Allen Telescope Array, etc.)
  are suitable to enable observing programs that are optimized for technosignature searches in 2020--2030.
  In 2018, BL used
  $\sim$1200 h
  of time on the 100~m Green Bank Telescope and $\sim$1500 h of time on the 64 m Parkes Telescope.
  These allocations increased the sampled fraction of the search volume (as quantified by the Drake figure of merit~\citep{drak84}) by a factor of several compared to the search of \citet{enri17}.  

\subsection{Additional results}
One type of search for technosignatures includes recording of the baseband, raw voltage data.  These data sets provide maximal flexibility in terms of signal processing and enable {\em de novo} analyses with the latest software developments long after the observations are conducted \citep[e.g.,][]{pinc19}.  Importantly, they enable the discovery of other astronomical sources or transients, such as pulsars or fast radio bursts (FRBs) \citep[e.g.,][]{gajj18}.  These additional results augment the astrophysical return on investment of telescope time.

\section{Key advances necessary for completion}

\subsection{Rebuilding of research ecosystem}

The research ecosystem in the field of technosignatures has been largely destroyed by the lack of federal funding in the past 25 years.  In order for this discipline to thrive, it is essential to restore the infrastructure and growth opportunities that are present in traditional scientific fields, including scientific workshops and conferences, university instruction and mentoring, fellowship programs for graduate students and postdocs, grant programs for data acquisition and analysis, major equipment programs, and representation on national committees that evaluate research priorities.

The commitment of federal dollars to the search for technosignatures is urgent.  The search for biosignatures has been enabled almost exclusively by federally-funded facilities and spacecraft (e.g., Hubble, Kepler, JWST, Mars 2020, Europa Clipper), without philanthropic or private sector partnerships.  The search for technosignature is an equally compelling scientific investigation that deserves federal funding, regardless of the availability of private funding.  

\subsection{Telescope resources}
The search requires both dedicated and commensal time on radio telescopes with large collecting areas.
Example facilities are listed in Section~\ref{sec-dedicated}.  Multibeam receivers are an important enabling technology because they increase sky coverage and survey speed by approximately an order of magnitude.  Examples include the ATA 3-beam wideband SonATA system, the Parkes 21cm Multibeam Receiver with 13 dual-polarization beams, and the planned ALPACA system
with 40 dual-polarization beams at Arecibo.  On telescope arrays, the availability of electronic beamforming is an important advance because it enables multiple pointings without slewing the individual elements, multiple simultaneous science goals, and commensal observations. 

\subsection{Hardware advances}

In order to maximize sampling of the cosmic haystack, radio telescopes should be outfitted with digital backends that sample the full receiver bandwidth.  Modular and programmable backends, such as the CASPER architecture \citep[e.g.,][]{hick16}, facilitate the implementation of on-board pre-processing, event triggering, and even search algorithms.  In situations where most of the raw data are not stored, the availability of memory ring buffers enable
storage of the relevant subset of raw voltage data based on 
near-real-time event detection.

\subsection{Software advances}      
Increases in survey bandwidths and speeds make data storage intractable and real-time processing of the data essential.  Example strategies are implemented in FRB searches~\citep[e.g.,][]{chim18}.  Another computational challenge is to excise or classify RFI.  Implementation of novel algorithms, including machine learning techniques
\citep[e.g.,][]{zhan19,harp19},
can improve signal classification performance.  Because of the large computational cost, GPU-accelerated and distributed-computing solutions will
be required.  In some situations, grid computing and citizen science initiatives may be appropriate.

\section{Conclusions}\label{sec:conclusions}
The radio search for technosignatures offers a compelling opportunity to address one of the most important questions in the history of science.  It enables an expansion of the search for life in the universe from primitive to complex life and from the solar neighborhood to the entire Galaxy.  With a modest investment of federal dollars, the field will
see dramatic advances in 2020--2030.

\newpage
\bibliography{radio}

\begin{thebibliography}{29}
\bibitem[{Ekers}(2002)]{eker02}
R.~D. {Ekers}.
\newblock \emph{{SETI 2020 : a roadmap for the search for extraterrestrial
  intelligence / produced for the SETI Institute by the SETI Science \&
  Technology Working Group}}.
\newblock SETI Press, 2002.
\end{thebibliography}

\end{document}